\documentclass[12pt]{article}
\usepackage{hhline}
\usepackage{latexsym,graphicx}
\usepackage{subfigure}
\usepackage{amssymb}
\usepackage{amsmath}
\usepackage{amscd}
\usepackage{amsthm}
\usepackage{float}
\usepackage[left=2cm,top=2.5cm,right=2.5cm,bottom=1.5cm]{geometry}   
\DeclareMathOperator{\dalembert}{\Box} 
\begin{document}
\begin{center}
\large{\bf{Ricci-Gauss-Bonnet holographic dark energy in Chern-Simons modified gravity: A flat FLRW quintessence-dominated universe}} \\
\vspace{10mm}
\normalsize{Nasr Ahmed$^1$}\\
\vspace{5mm}
\small{\footnotesize $^1$ Astronomy Department, National Research Institute of Astronomy and Geophysics, Helwan, Cairo, Egypt\footnote{abualansar@gmail.com}} 
\end{center}  
\date{}
\begin{abstract}
We discuss the recently suggested Ricci-Gauss-Bonnet holographic dark energy in Chern-Simons modified gravity. We have tested some general forms of the scale factor $a(t)$, and used two physically reasonable forms which have been proved to be consistent with observations. Both solutions predict a sign flipping in the evolution of cosmic pressure which is positive during the early-time deceleration and negative during the late-time acceleration. This sign flipping in the evolution of cosmic pressure helps in explaining the cosmic deceleration-acceleration transition, and it has appeared in other cosmological models in different contexts. However, the current work shows a pressure singularity which needs to be explained. The evolution of the equation of state parameter $\omega(t)$ shows the same asymptotic behavior for both solutions indicating a quintessence-dominated universe in the far future. We also note that $\omega(t)$ goes to negative values (leaving the decelerating dust-dominated era at $\omega=0$) at exactly the same time the pressure becomes negative. Again, there is another singularity in the behavior of $\omega(t)$ which happens at the same cosmic time of the pressure singularity.

\end{abstract}

\section{Introduction and motivation}

A major challenge to the current understanding of the standard models of gravity is the discovery of the late-time accelerating exapansion of the universe \cite{11,13,14}. It has also been indicated by observations that the universe is flat, highly homogeneous and isotropic on large scales \cite{teg,ben,sp}. On the other hand, dark energy is an exotic form of energy with negative pressure that has been suggested to explain this cosmic acceleration. Several dark energy models have been proposed in the literature through modified gravity theories \cite{quint}-\cite{nass} and dynamical scalar fields \cite{1,moddd,39}, \cite{noj1}-\cite{alt3}. Examples of modified gravity theories include $f(R)$ gravity \cite{39} where $R$ is the Ricci scalar, Gauss-Bonnet gravity \cite{noj8} and f(T) gravity \cite{torsion} where $T$ is the torsion scalar. While $f(R)$ gravity replaces $R$ in the Einstein-Hilbert action by an arbitrary function $f(R)$, the fundamental ingredient in $f(T)$ gravity is the torsion scalar $T$ instead of $R$. Gauss-Bonnet gravity replaces $R$ in the Einstein-Hilbert action by the Gauss-Bonnet term $G=R^2-4R^{\mu\nu}R_{\mu\nu}+R^{\mu\nu\rho\delta}R_{\mu\nu\rho\delta}$, a general function $f(G)$ can be used in the action instead of $G$. A generalization of $f(R)$ gravity has been done through introducing an arbitrary function $f(R,T)$ \cite{1}, where $T$ is the trace of the energy momentum tensor. \par
A particular modified gravity theory that has gained a remarkable attention in recent years is Chern-Simons modified gravity \cite{cs}. This modified gravity theory is an extension to general relativity GR aimed to solve the long standing problem of cosmic baryon asymmetry by introducing a parity-violating term to the Einstein-Hilbert action. The gravitational field in this modified gravity is coupled with a scalar field using a parity-violating Chern-Simons term. Adding the gravitational Chern-Simons term to the Einstein-Hilbert action violates the Lorentz and CPT symmetries \cite{cs, cs11,cs22,cs33}. It has also been shown that this term arises as a quantum correction in a theory describing coupling of gravity to fermions \cite{cs11,cs22,cs33}, such a correction is unavoidable in string theory to remain mathematically consistent \cite{cs44,cs55}. The Chern-Simons term is defined as a contraction of the Riemann curvature tensor with its dual and the Chern-Simon scalar field \cite{cs}. Cosmological solutions of Chern-Simons modified gravity with several holographic dark energy models have been discussed in \cite{cs1, cs2, cs3, cs4}. \par
Based on the holographic principle in quantum gravity \cite{holo}, the holographic dark energy model has been constructed in \cite{hde}. This model links the dark energy density to the cosmic horizon \cite{hde1, hde2} and has been tested through observations \cite{hde3}. The Ricci Dark Energy model is inversely proportional to Ricci scalar curvature and has been introduced in \cite{rde}. A new model of holographic dark energy has been presented in \cite{gbhde} in which the Infrared cutoff is determined by both the Ricci and the Gauss-Bonnet invariants. A major advantage of this new model is that the Infrared cutoff is determined by invariants which play a fundamental role in theories of gravity. In this new scenario of holographic dark energy, the inverse squared IR cutoff is given by $\frac{1}{L^2}=-\alpha R+\beta \sqrt{\left|G\right|}$ where $\alpha$ and $\beta$ are constants. The standard Ricci dark energy can be obtained for $\beta=0$, while a pure Gauss-Bonnet holographic dark energy can be obtained for $\alpha=0$. Cosmological analytical solutions have been discussed in \cite{gbhde} focusing on the matter-dominated case for simplicity. The equation of state parameter of the Ricci-Gauss-Bonnet holographic dark energy can lie in the quintessence or phantom regime, or cross the phantom-divide line. Its asymptotic value in the far future depends on the model parameters and can be quintessence-like, phantom-like, or be equal to the cosmological constant \cite{gbhde}.\par
The current work discusses the evolution of Ricci-Gauss-Bonnet holographic dark energy in the context of Chern-Simons modified gravity which has not been discussed before in the literature. We have adopted an empirical approach to solve the cosmological equations so that the behavior of the deceleration and jerk parameters is consistent with observations and the flat standard cosmological model-$\Lambda$CDM. Another advantage of the current work is that we have obtained a positive-to-negative transition in the behavior of cosmic pressure which is so helpful in explaining the deceleration-to-acceleration cosmic transition. Finally, it's interesting to obtain the same asymptotic value in the far future for two different empirical solutions describing a future quintessence-dominated universe. The paper is organized as follows: In section 2, we review the Ricci-Gauss-Bonnet Holographic dark energy and the field equations of Chern-Simons modified gravity. The cosmological solutions have been discussed in section 3. The final conclusion is included in section 4.
\section{Ricci-Gauss-Bonnet HDE in CS Modified Gravity} \label{sol}
The action of Chern-Simon modified gravity is given by
\begin{eqnarray}
S&=&S_{EH}+S_{CS}+S_{\theta}+S_{mat},\\    \nonumber
&=&\kappa \int{d^{4}x \sqrt{-g}~R}+\frac{\alpha}{4}\int{d^{4}x \sqrt{-g}~\theta~ ^{*}RR}-\frac{\beta}{2}\int{d^{4}x \sqrt{-g}\left[\partial^{\mu}\theta \partial_{\mu}\theta +2V(\theta)\right]}+S_{mat},
\end{eqnarray}
The CS coupling field $\theta$ is serving as a deformation function of space-time, CS modified gravity reduces to GR for $\theta=0$. The matter contribution are described by $S_{mat}=\int{d^{4}x \sqrt{-g} L_{mat}}$ where $L_{mat}$ doesn't depend on the scalar field $\theta$. $\kappa^{-1}=16\pi G$, $\alpha$ and $\beta$ are dimensionless constants. The quantity $^{*}RR$ is the Pontryagin density defined as $^{*}RR=^{*}R^{a~cd}_{~b}R^b_{~acd}$ where the dual Riemann-tensor is given by $R^{a~cd}_{~b}=\frac{1}{2}\epsilon^{cdef}R^a_{~bef}$ with $\epsilon^{cdef}$ is the 4-dimensional Levi-Civita tensor.  The potential $V(\theta)$ is set to zero in the current work for simplicity. Varying the action with respect to the metric $g_{\mu\nu}$ and to the scalar field $\theta$ we get the field equations as
\begin{equation}\label{fe}
G_{ab}+\frac{\alpha}{\kappa}C_{ab}=\frac{1}{2\kappa}T_{ab},
\end{equation}
\begin{equation} \label{th}
\beta \dalembert \theta=-\frac{\alpha}{4} ^{*}RR.
\end{equation}
Where $G_{\mu\nu}$ is the Einstein tensor and $C_{\mu\nu}$ is the Cotton tensor. The
energy-momentum tensor is composed of two parts, the matter part $T_{\mu\nu}^m$ and the scalar field part $T_{\mu\nu}^{\theta}$ where
\begin{equation}
T^m_{\mu\nu}=(\rho+p)u_{\mu}u_{\nu}-pg_{\mu\nu},
\end{equation}
\begin{equation}
T^{\theta}_{\mu\nu}= \nabla_{\mu} \theta  \nabla_{\nu} \theta -\frac{1}{2}g_{\mu\nu} \nabla^{\lambda} \theta  \nabla_{\lambda} \theta,
\end{equation}
Where $\rho$ is energy density, $p$ is pressure and $u$ is the four velocity vector. The energy density of the recently suggested Ricci-Gauss-Bonnet holographic dark energy is given by \cite{gbhde},
\begin{equation}
\rho_D=\frac{3}{\kappa^2}\left[ 6\alpha(2H^2+\dot{H})+2\sqrt{3}\beta H \sqrt{\left|H^2+\dot{H}\right|}\right]
\end{equation}
We consider the flat FLRW metric, given by
\begin{equation} \label{frw}
ds^{2}=-dt^{2}+a^{2}(t)\left[ dr^{2}+r^2d\theta^2+r^2\sin^2\theta d\phi^2 \right] 
\end{equation} 
where $a(t)$ is the cosmic scale factor, $H=\frac{\dot{a}}{a}$ is the Hubble parameter, and the dot denotes the time derivative. Using the metric (\ref{frw}), the 00-component of equation (\ref{fe}) leads to the following Friedmann equation 
\begin{equation} \label{fre}
H^2=\frac{\kappa^2}{3} \rho_D+\frac{1}{6}\dot{\theta}^2,
\end{equation}
The value of $\theta$ can be calculated using equation (\ref{th}). The term $^{*}RR=^{*}R^{a~cd}_{~b}R^b_{~acd}$ vanishes for FLRW universe and then equation (\ref{th}) becomes
\begin{equation} 
\beta \dalembert \theta=0
\end{equation}
which gives
\begin{equation} 
\dot{\theta}=Ca^{-3},
\end{equation}
where $C$ is a constant of integration. Substituting in (\ref{fre}), we get for the flate universe
\begin{equation} 
H^2=\left[ 6\alpha(2H^2+\dot{H})+2\sqrt{3}\beta H \sqrt{\left|H^2+\dot{H}\right|}\right]+\frac{1}{6} (Ca^{-3})^2
\end{equation}
Using the conservation equation
\begin{equation} 
\dot{\rho}_{\Lambda}+3H(\rho_{\Lambda}+p_{\Lambda})=0
\end{equation}
Then, the pressure $p_{\Lambda}$ can be written as
\begin{equation} 
p_{\Lambda}=-\frac{1}{H} \frac{d}{dt}\left[ 6\alpha(2H^2+\dot{H})+2\sqrt{3}\beta H \sqrt{\left|H^2+\dot{H}\right|}\right]-6\alpha (2H^2+\dot{H})-2\sqrt{3}\beta H \sqrt{\left|H^2+\dot{H}\right|}
\end{equation}
And the equation of state parameter $\omega_{\Lambda}=\frac{p_{\Lambda}}{\rho_{\Lambda}}$ can be expressed as 
\begin{equation} 
\omega_{\Lambda}= \frac{F(t)}{G(t)}
\end{equation}
Where
\begin{eqnarray} 
F(t)&=&\left(\ddot{H}+\alpha H(H^2+\frac{5}{2}\dot{H}) \sqrt{\left|H^2+\dot{H}\right|}\right)\\   \nonumber
&+&\frac{\sqrt{3}}{3}\beta \left(H(H\dot{H}+\frac{1}{2}\ddot{H})\left(\sqrt{\left|H^2+\dot{H}\right|}\right)^{-1}+(H^2+\dot{H})\left|H^2+\dot{H}\right|\right)
\end{eqnarray}
and
\begin{equation} 
G(t)=-\left(\sqrt{\left|H^2+\dot{H}\right|}H\left(\sqrt{3}\beta H \sqrt{\left|H^2+\dot{H}\right|}+6\alpha(H^2+\frac{1}{2}\dot{H})\right)\right)^{-1}
\end{equation}
\section{Cosmological solutions}
Some general scale factor ansatze have been investigated in \cite{basic11} in order to construct bouncing solutions in modified Gauss-Bonnet gravity $f(G)$, with $G$ the Gauss-Bonnet invariant. They have investigated the case where the scale factor has the form of linear combination of two exponential terms $a(t)=\sigma e^{\lambda t}+\tau e^{-\lambda t}$, the case for $a(t)=\frac{1}{2}(e^{\lambda t}+ e^{-\lambda t})=\cosh(\lambda t)$, the exponential case $a(t)=e^{\alpha t^2}$, the power-law case $a(t)=\beta t^{2n}$, and the case for the sum of multiple exponential functions $a(t)=e^{\alpha t^2}+e^{\alpha^2 t^4}$. Another interesting ansatz has been introduced in the so called logamediate inflation scenario \cite{logscale} where the scale factor expands as $a(t)=e^{A\ln(t)^{\lambda}}$, and it has been proved to be consistent with CMB observations. In order to find a solution which is consistent with observations, we have investigated all these ansatze according to the observationally suggested cosmic transition \cite{11,rrr}, and the physically accepted positivity of energy density. Table 1 shows that, for the current model, only two forms are physically reasonable and satisfying the observational requirement of the cosmic deceleration-to-acceleration transition: the hyperbolic form $a(t)= A\sinh^r(\xi t)$ with $0 < r < 1$, and the logamediate inflation form $a(t)=e^{A\ln(t)^{\lambda}}$. The ansatz $a(t)=\sigma e^{\lambda t}+\tau e^{-\lambda t}$ gives a physically acceptable behavior for the energy density but no cosmic transition. The three forms $a(t)=\beta \cosh(\lambda t)$, $a(t)=e^{\alpha t^2}+e^{\alpha^2 t^4}$, and $a(t)=e^{\alpha t^2}$ lead to a wrong behavior for the energy density where $\rho(t) \rightarrow 0$ as $t \rightarrow 0$. Based on this analysis, we consider the two forms  satisfying both conditions in table 1 as empirical forms, and use them to explore possible solutions. The following hyperbolic form produces a good agreement with observations for $0 < r < 1$
\begin{equation} \label{ansatz}
a(t)= A\sinh^r(\xi t)
\end{equation}
For $r=\frac{1}{2}$ and $A=\xi=1$, this gives the time-varying deceleration and jerk parameters respectively as
\begin{equation} \label{q1}
q(t)=-\frac{\ddot{a}a}{\dot{a}^2}=\frac{-\cosh^2(t)+2}{\cosh^2(t)}
\end{equation}
\begin{equation}
j(t)=\frac{\dddot{a}a^2}{\dot{a}^3}=\frac{\cosh^2(t)+2}{\cosh^2(t)}
\end{equation}
The jerk parameter provides a convenient method to describe models close to $\Lambda$CDM \cite{jerk1,jerk2}, flat $\Lambda$CDM models have $j = 1$ \cite{81}. Figure \ref{F6202y28} shows that for the current flat model the jerk parameter has the asymptotic value $j=1$ at late-time. So, in addition to the deceleration-to-acceleration transition behavior of the deceleration parameter $q(t)$, the behavior of the jerk parameter $j(t)$ at the current epoch represents another support for using this hyperbolic ansatz. Such hyperbolic form appears in several cosmological models. It has been used in the study of Bianchi cosmological models where a good agreement with observations has been obtained \cite{pr}. A Quintessence model with double exponential potential has been constructed in \cite{sen} assuming the form $a(t)= \frac{a_o}{\alpha}[\sinh(t/t_o)]^{\beta}$, where $t_o$ is the present time, $R_o$ is the present day scale factor, $\beta$ is a constant and $\alpha = [\sinh(1)]^{\beta}$. As has been mentioned in \cite{sen}, the main motivation for assuming this form is its consistency with observations as it gives both early-time deceleration and late-time acceleration. It has been shown in \cite{senta} that a solution of the form $a(t)= A ~(\sinh \sqrt{2}\nu (t+t_{pl}))^{\frac{1}{2}}$ represents an exact solution of scalar field cosmology in modified $f(R)$ gravity. In the study of Ricci dark energy in Chern-Simons modified gravity, it has been shown that the evolution of the scale factor is given by $a(t)=\left(\frac{2\zeta}{3c_1}\right)^{\frac{1}{6}} \sinh^{\frac{1}{3}}(3\sqrt{c_1t})$ \cite{sent}, where $\zeta$ and $c_1$ are constants. In the context of $\Lambda$CDM model, such hyperbolic solution has been obtained describing the cosmic evolution from the matter-dominated epoch up to the late-time future \cite{sz}. 
\begin{table}\label{tap}
\centering
\tiny
    \begin{tabular}{ | p{3.6cm} | p{2.5cm} | p{2.5cm} |}
    \hline
     $a(t)$ ansatz & $q(t)$ sign flipping \newline from +ve to -ve & $\rho(t) \rightarrow \infty$ as $t \rightarrow 0$ \\ \hline
   $a(t)= A\sinh^r(\xi t)$, ~ $0 < r < 1$ & $\checkmark $ & $\checkmark$  \\ \hline
   $\sigma e^{\lambda t}+\tau e^{-\lambda t}$ & $\times$ (Ever accelerating) & $\checkmark$    \\ \hline
    $\beta \cosh(\lambda t)$ & $\times$ (Ever accelerating) & $\times$ ($\rho(t) \rightarrow 0$ as $t \rightarrow 0$)\\ \hline
	$e^{\alpha t^2}$ & $\checkmark$ & $\times$ ($\rho(t) \rightarrow 0$ as $t \rightarrow 0$) \\ \hline
		$\beta t^{2n}$ & $\times$ (constant) & $\checkmark$ for $n<0$.  \\ \hline
		$e^{\alpha t^2}+e^{\alpha^2 t^4}$ & $\checkmark $ with singularity& $\times$ ($\rho(t) \rightarrow 0$ as $t \rightarrow 0$)\\ \hline
		$e^{A \ln(t)^{\lambda}}$ & $\checkmark $ & $\checkmark $\\ \hline
    \end{tabular}
		\caption {For $a(t)=\sigma e^{\lambda t}+\tau e^{-\lambda t}$ and $a(t)=\beta \cosh(\lambda t)$, $q$ is always negative but tends to $-1$ at late time in a good agreement with observations. The two cases of $a(t)=e^{\alpha t^2}$ and $a(t)=e^{\alpha t^2}+e^{\alpha^2 t^4}$ are not physically acceptable for the current model. $a(t)=(\sinh(\xi t))^{\frac{1}{2}}$ and $a(t)=e^{A \ln(t)^{\lambda}}$ are physically acceptable forms and allow a cosmic transition.}
		\end{table}
\subsection{The hyperbolic solution.}
\begin{figure}[H]
  \centering            
  \subfigure[$q$]{\label{F63}\includegraphics[width=0.3\textwidth]{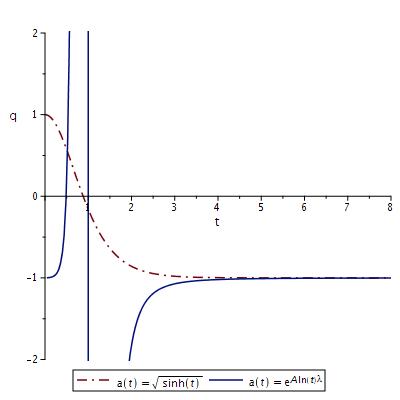}} 
\subfigure[$j$]{\label{F6202y28}\includegraphics[width=0.3\textwidth]{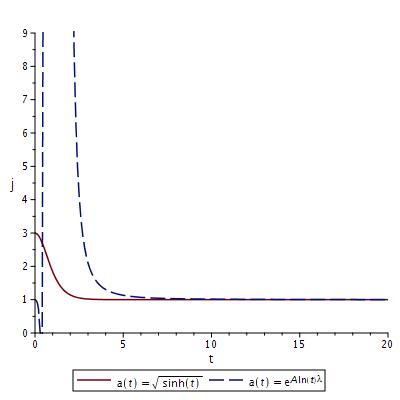}}
\subfigure[$a$]{\label{F622}\includegraphics[width=0.3\textwidth]{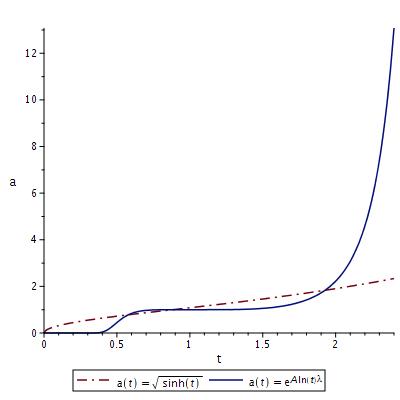}} \\
 \subfigure[$p_{\Lambda}$ for $\sinh^{\frac{1}{2}}(t)$]{\label{F6222}\includegraphics[width=0.3\textwidth]{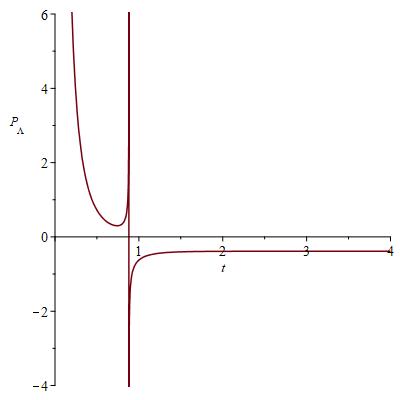}}
 \subfigure[$\rho_{\Lambda}$ for $\sinh^{\frac{1}{2}}(t)$]{\label{F62262}\includegraphics[width=0.3\textwidth]{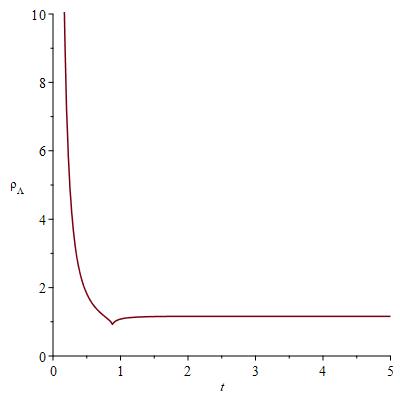}}
 \subfigure[$\omega_{\Lambda}$ for $\sinh^{\frac{1}{2}}(t)$]{\label{F62272}\includegraphics[width=0.3\textwidth]{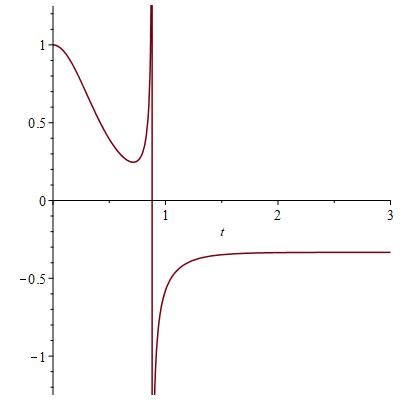}}\\
\subfigure[$p_{\Lambda}$ for$e^{A \ln(t)^{\lambda}}$]{\label{F62228}\includegraphics[width=0.3\textwidth]{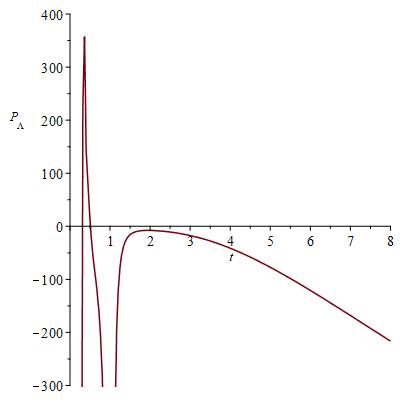}}
\subfigure[$\rho_{\Lambda}$ for$e^{A \ln(t)^{\lambda}}$]{\label{F622u28}\includegraphics[width=0.3\textwidth]{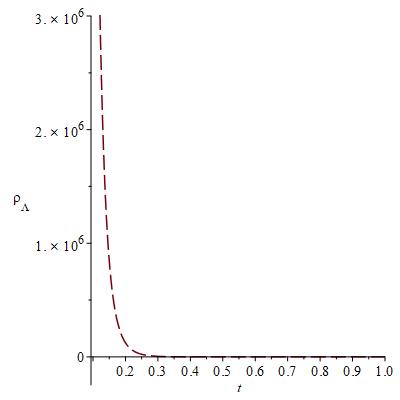}}
\subfigure[$\omega_{\Lambda}$ for$e^{A \ln(t)^{\lambda}}$]{\label{F622y28}\includegraphics[width=0.3\textwidth]{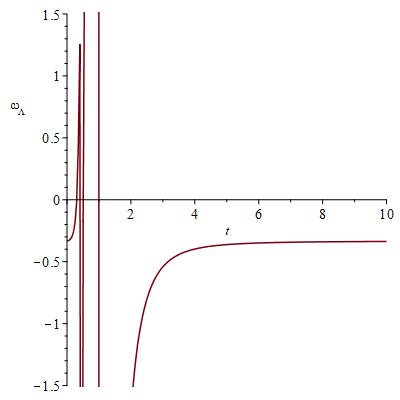}}
\caption{\ref{F63} and  \ref{F6202y28}: The deceleration and jerk parameters for both solutions. \ref{F622} the scale factor for both solutions. \ref{F6222}, \ref{F62262} and \ref{F62272} : The pressure, energy density and EoS parameter for the hyperbolic solution. 
\ref{F62228}, \ref{F622u28}, \ref{F622y28}: The pressure, energy density, EoS parameter for the second solution solution. Here $\alpha=\beta=0.1$}
  \label{fig:cassimir55}
\end{figure}

In this case, we get
\begin{eqnarray}
p(t)&=&\frac{1}{\sqrt{f}g^3h}\left(\frac{\sqrt{3}}{4}\beta \coth^2(t)\frac{(2-h^2)\left|4 g^2\right|}{\left| 2-h^2\right|}-3\alpha g^3 h\sqrt{f}(h^2+1)
+\sqrt{3}\beta f (h^2-2)\right),\\ \nonumber
\rho_{\Lambda}(t)&=&\frac{3}{2g}\left(\sqrt{3} \beta h\sqrt{f}+6\alpha g \right), \\  \nonumber
\omega(t)&=&\frac{2\left(\frac{\sqrt{3}}{4}\beta \coth^2(t)\frac{(2-h^2)\left|4 g^2\right|}{\left| 2-h^2\right|}-3\alpha g^3 h\sqrt{f}(h^2+1)
+\sqrt{3}\beta f (h^2-2)\right)}{3hg^2\sqrt{f} \left(\sqrt{3} \beta h\sqrt{f}+6\alpha g \right)}.
\end{eqnarray}
where $f=\left| \frac{\cosh^2(t)-2}{\sinh^2(t)}\right|$, $g=\sinh(t)$ and $h=\cosh(t)$. Figure \ref{F63} shows the sign flipping of the deceleration parameter (\ref{q1}) from positive (decelerating phase) to negative (accelerating phase) for $r=\frac{1}{2}$. It varies in the range $-1\leq q \leq 1$, the starting value is $q=1$ represents a decelerating radiation-dominated
era which agrees with the complete cosmic history described in \cite{cosmich}. Then, it passes the matter-dominated era at $q=\frac{1}{2}$ and ends at $q=-1$ which represents an accelerating era. The behavior of the cosmic pressure is illustrated in Figure \ref{F6222}. The pressure is positive during the early radiation and matter-dominated time where the expansion was decelerating, and negative during the late dark energy-dominated time where the expansion is accelerating. This behavior agrees with the standard cosmological model where the early decelerating  universe ($z\rightarrow \infty$) is filled with positive pressure \cite{dd}. The negative pressure which dominates the late-time universe represents the anti-gravity effect that pushes the universe to expand faster and faster. This has been also shown in the context of the causal Israel-Stewart formalism where a positive pressure with viscosity leads to a decelerating expansion \cite{ddd}. \par
Investigating the evolution of the EoS parameter $\omega = \frac{p}{\rho}$ and detecting its current value is an important step in understanding the nature of dark energy. This parameter is equal to $0$ for dust, $1/3$ for radiation, $-1$ for vacuum energy, $ \leq -1$ for phantom scalar field. We also have $-1 \leq \omega \leq 1$ for quintessence scalar field, and it can evolve across the cosmological constant boundary $\omega = -1$ for quintom field. Quintom is a dynamical model of dark energy differs from cosmological constant, Quintessence, Phantom ...etc. by an important feature where the EoS parameter can smoothly cross over $\omega=-1$. The EoS parameter $\omega=1$ for some exotic type of matter called stiff matter \cite{zeld} where the speed of sound is equal to the speed of light, and it is the largest value of $\omega$ consistent with causality. The evolution of the EoS parameter $\omega(t)$ is shown in Figure \ref{F62272}. It starts at $\omega=1$ (stiff matter-dominated era), and keeps decreasing passing the radiation-dominated era at $\frac{1}{3}$ and dust-dominated era at $0$. It then goes to negative values (leaving the decelerating dust-dominated era at $0$) at exactly the same time the pressure becomes negative (see Fig. \ref{F6222}). Also, there is a pressure singularity and another singularity in the evolution of $\omega(t)$. In both of them, the singularity happens at the same time $t \approx 0.85$. A Quintom behavior exists in a narrow range of cosmic time where there is a crossing to the phantom divide line $\omega=-1$. Then, the evolution continues to a future Quintessence-dominated universe with a constant value of $\approx -\frac{1}{3}$. So, the present scenario is a quintessence-dominated universe with negative pressure.
\subsection{Logamediate Inflation $e^{A \ln(t)^{\lambda}}$.}
In this case, we get
\begin{eqnarray}
j(t)&=&\frac{1}{A^2\lambda^2}\left[(-3\lambda+3)l^{-2\lambda+1}+2l^{-2\lambda+2}-3A\lambda l^{-\lambda+1} \right. \\ \nonumber
&+& \left.(\lambda^2+3\lambda+2)l^{-2\lambda}+(3A\lambda^2-3A\lambda)l^{-\lambda}+A^2\lambda^2\right].
\end{eqnarray}

\begin{equation}
q(t)={\frac { \left( -\lambda+1 \right) l^{-\lambda}-A\lambda+ l^
{-\lambda+1}}{A\lambda}}~~~~~~~~~~~~~~~~~~~~~~~~~~~~~~~~~~
\end{equation}

\begin{eqnarray}\label{p}
p(t)=\frac {-2}{\sqrt { \left| \lambda \right| }\sqrt { \left| -
  l^{\lambda+1}+ l^{2\,\lambda}A\lambda+ \left( \lambda-1
 \right)  l^{\lambda} \right| }
\sqrt { \left| A \right| }{t}^{3} l^{3} \left| t \right| }\times~~~~~~~~\\   \nonumber
\beta\,\lambda\, \left| l \right| \sqrt {3}A
 \left(  \left| t \right|  \right) ^{2} \left(-A l^{2\,\lambda+1}\lambda+
 \frac{3}{2}( 1-\lambda)  l^{\lambda+1}+ l^{\lambda
+2}
+ \left( A{\lambda}^{2}-A\lambda \right) l^{2\,\lambda} \right.\\   \nonumber 
\left. +  \left|{t}^{2}\right| \frac{\left( \frac{1}{2}\,{\lambda}^{2}-\frac{3}{2}\,\lambda+1 \right)  l^{\lambda}A\lambda\, \left( l^{2\,
\lambda-2}A\lambda+ \left( \lambda-1 \right)  l^{\lambda-2}- l^{\lambda-1} \right) }{t^2\left|{\lambda\, \left( l^{2\,
\lambda-2}A\lambda+ \left( \lambda-1 \right) l^{\lambda-2}- l^{\lambda-1} \right) A}\right|}\right. \\   \nonumber 
\left. +6t\left(\,\beta\, \left| l \right|  \left( A l^{\lambda+2}\lambda+ l^{2} \left( \lambda-l -
1 \right)\sqrt {3} \left| \lambda \right| t \left| A \right| \times\right.\right.\right. ~~~~~~~~~~~~~~\\   \nonumber 
\left.\left.\left. \left|  l^{2\,\lambda-2}A\lambda
+ \left( \lambda-1 \right)  l^{\lambda-2}- l^{\lambda-1}
 \right| + \left({A}^{2} l^{2\,\lambda+1}{\lambda}^{2}+\frac{5}{2}\,A\lambda\, \left( \lambda-1 \right) l^{\lambda+1}\right.\right.\right.\right. ~~~~~\\   \nonumber 
\left.\left.\left.\left.
-\frac{5}{2}\,A l^{\lambda+2}\lambda+\frac{1}{2}\,l\left( 2\,
 l^{2}+ \left( -3\,\lambda+3\right) l+{\lambda}^{2}-3\,\lambda+2 \right) \right)\times  \right.\right.\right. ~~~~~\\   \nonumber 
\left.\left.\left. \alpha\, \left| t \right| \sqrt { \left| \lambda \right| }\sqrt {
 \left| - l^{\lambda+1}+ l^{2\,\lambda}A\lambda+ \left( \lambda-1
 \right) l^{\lambda} \right| } \sqrt { \left| A \right| }
\right)
\right)
\right)~~~~~~~~~~~~
\end{eqnarray}

\begin{eqnarray} \label{rho}
\rho(t)&=&\frac{6A\lambda}{t^2\left|t\right|}\times \left(6\, l^{2\,\lambda-2}A\alpha\,\lambda\, \left| t \right| +\right.   \\   \nonumber
&&\left. \left( \beta\,t\sqrt {3} \left| \sqrt {A\lambda{\frac {(- l^{\lambda+1}+ l^{2\,\lambda}A
\lambda+ \left( \lambda-1 \right) l^{\lambda})}{ l^{2}}}}
 \right| -3\,\alpha\, \left| t \right|  \right)  l^{\lambda-1}\right.  \\  \nonumber
&&\left.+3\, l^{\lambda-2}\alpha\, \left| t \right|  \left( \lambda-1
 \right)
\right)
\end{eqnarray}
where $l=\ln(t)$. The expression for the EoS parameter is directly obtained from (\ref{p}) and (\ref{rho}) as $\omega(t)=p(t)/\rho(t)$. Figure \ref{F63} shows the evolution of the deceleration parameter for the Logamediate Inflation ansatz. The sign changes from negative (at the very early universe) to positive, and then comes back to negative again. So, this ansatz gives a more general description than the previous hyperbolic one. It includes both the acceleration-to-deceleration transition which is important in the very early universe, and the late-time deceleration-to-acceleration transition which is important in the current epoch. A model that includes both very early and late-time transitions can provide a better explanation to all observational phenomena and a deeper understanding to the whole picture of cosmic evolution. The behavior of this ansatz with cosmic time in Figure \ref{F622} shows a pre-radiation era, an early inflation era, the radiation era and a late-inflation era successively (see \cite{vans} for a description of a similar scale factor). Figure \ref{F62228} shows that the pressure evolves from negative (at the very early universe) to positive, and then to negative again at the current epoch. This behavior agrees with the deceleration parameter evolution (Figure \ref{F63}) where positive pressure represents the attractive gravity associated with the decelerating expansion and the negative pressure represents the  repulsive gravity associated with the accelerating expansion. \par
The EoS parameter of this solution has the asymptotic value $\approx -\frac{1}{3}$ which is exactly the same asymptotic value of the Eos parameter of the hyperbolic solution. This is interesting as both solutions predict the same future quintessence-dominated universe with the same asymptotic value. It has been shown in \cite{gbhde} that the EoS parameter of the Ricci-Gauss-Bonnet holographic dark energy can lie in the quintessence or phantom regime, or cross the phantom-divide line. It has also been indicated that the asymptotic value of the EoS parameter in the far future depends on the model parameters and can be quintessence-like, phantom-like, or be equal to the cosmological constant. Like the case of the hyperbolic solution, there is also a singularity in $p(t)$ and $\omega(t)$ for the current solution. It is also interesting to note that both singularities happen at the same cosmic time of the hyperbolic solution $t\approx 0.85$ at which the sign-flipping of $p(t)$ and $\omega(t)$ happens. Again, the present scenario is a quintessence-dominated universe with negative pressure as predicted by the first solution.

\section{Conclusion}

The recently suggested Ricci-Gauss-Bonnet holographic dark energy has been investigated in Chern-Simons modified gravity through two physically reasonable scale factor forms. The empirical hyperbolic ansatz $a(t)= \sinh^{r}(\xi t)$ for $0 < r < 1$ results in a cosmic pressure which is positive during the early-time deceleration and negative during the late-time acceleration. It also predicts a universe that is dominated by a quintessence field in the far future with asymptotic value of the EoS parameter $\omega=-\frac{1}{3}$. The another solution is the Logamediate inflation $a(t)=e^{A \ln(t)^{\lambda}}$ which has been proved to be consistent with observations. This form resulted in a cosmic pressure that turns from negative (at the very early universe) to positive and then to negative again (at the current era). It is interesting to find the following similarities between the two different solutions:
\begin{itemize}
\item Both solutions predicting a sign-flipping from negative to positive in the evolution of $p(t)$, the behavior which helps in explaining the observationally suggested cosmic transit from deceleration to acceleration. Such negative-to-positive transition in the evolution of cosmic pressure appears in different cosmological contexts such as cyclic universes \cite{cyc}, entropy-corrected cosmology \cite{ent1,ent2}, and Swiss-cheese brane-world cosmology \cite{br1}. The second solution gives a more general description to the evolution of cosmic pressure than the first one. 
\item The evolution of both $p(t)$ and $\omega(t)$ has a singularity which happens at the same cosmic time in both of them $t \approx 0.85$. $\omega(t)$ leaves the decelerating dust-dominated era at $\omega=0$ and crosses to negative values at exactly the same time where the cosmic pressure becomes negative.
\item Both solutions reveal the same asymptotic behavior of the $\omega(t)$ predicting a quintessence-dominated universe in the far future at the same asymptotic value $\omega=-\frac{1}{3}$. 
\item The present scenario in both solutions is that we are living in a flat FLRW accelerating and quintessence-dominated universe with negative pressure.
\end{itemize}

\end{document}